\documentclass[twoside,english,11pt]{article}
\usepackage[T1]{fontenc}
\usepackage[latin9]{inputenc}
\usepackage{geometry}
\usepackage{amsmath}
\usepackage{amsthm}
\usepackage{amssymb}
\usepackage{amstext}
\usepackage{bm}
\usepackage{booktabs} 
\usepackage{color}
\usepackage{enumerate} 
\usepackage[shortlabels]{enumitem}
\usepackage{epsfig,epsf,psfrag}
\usepackage{epstopdf}
\usepackage{float}
\usepackage{graphics}
\usepackage{graphicx}
\usepackage{latexsym}
\usepackage{lscape}
\usepackage{mathabx} 
\usepackage{mathtools}
\usepackage{mathrsfs}
\usepackage{multirow}
\usepackage{natbib}
\usepackage{rotate}
\usepackage{setspace}

\usepackage{sectsty}

\usepackage{subcaption}
\usepackage{xargs}[2008/03/08]
\usepackage{xcolor}
\usepackage{xspace}

\def\update{}
\def\revtwo{}

\definecolor{dgray}{gray}{0.35}

\def\note#1{\bco#1\fi}

\usepackage{ifthen}
\usepackage{fancyhdr}
\usepackage[hidelinks]{hyperref}
\hypersetup{
	colorlinks,
	linkcolor={red!50!black},
	citecolor={blue!50!black},
	urlcolor={blue!80!black}
}

\renewcommand\thesection{\arabic{section}}
\renewcommand{\thesubsection}{\thesection.\arabic{subsection}}

\sectionfont{\centering\bf\large}
\subsectionfont{\normalsize}

\newcommand{\Appendix}{\appendix
	\def\thesection{Appendix:}
	\def\thesubsection{\Alph{section}.\arabic{subsection}}}

\definecolor{DarkRed}{rgb}{.7,0,.4}

\renewcommand{\baselinestretch}{1.6} 
\newcommand{\single}{\renewcommand{\baselinestretch}{1.2}\normalsize}

\renewenvironment{thebibliography}[1]{
	\begin{oldthebibliography}{#1}
		\setlength{\parskip}{0ex}
		\setlength{\itemsep}{0ex}
	}
	{
	\end{oldthebibliography}
}

\newtheorem{thm}{Theorem}
\newtheorem{lem}{Lemma}
\newtheorem{cor}{Corollary}
{
	\theoremstyle{remark}
	\newtheorem{rem}{Remark}
	\newtheorem{eg}{Example}
	\newtheorem{assumption}{Assumption}
}

\newtheorem*{lem*}{Lemma}

\newcommand{\bea}{\begin{eqnarray*}}
\newcommand{\eea}{\end{eqnarray*}}
\newcommand{\be}{\begin{eqnarray}}
\newcommand{\ee}{\end{eqnarray}}
\newcommand{\beq}{\begin{equation}}
\newcommand{\eeq}{\end{equation}}

\newcommand{\bal}{\begin{equation}\aligned}
\newcommand{\eal}{\endaligned\end{equation}}
\newcommand{\bgt}{\begin{equation}\begin{gathered}}
\newcommand{\egt}{\end{gathered}\end{equation}}
\newcommand{\ed}{\end{document}}

\newcommand{\btab}{\begin{tabular}}
\newcommand{\etab}{\end{tabular}}

\newcommand{\bi}{\begin{itemize}}
\newcommand{\ei}{\end{itemize}}
\newcommand{\bfi}{\begin{figure}}
\newcommand{\efi}{\end{figure}}
\newcommand{\ben}{\begin{enumerate}}
\newcommand{\een}{\end{enumerate}}
\newcommand{\bay}{\begin{array}}
\newcommand{\eay}{\end{array}}

\newcommand{\nn}{\nonumber}

\def\bco{\iffalse}

\newcommand{\no}{\noindent}
\newcommand{\bc}{\begin{center}}
\newcommand{\ec}{\end{center}}

\definecolor{DarkBlue}{rgb}{0,.08,.45}
\definecolor{DarkRed}{rgb}{.7,0,.4}

\newcommand{\bpmt}{\begin{pmatrix}}
	\newcommand{\epmt}{\end{pmatrix}}

\newcommand{\bsp}{\begin{split}}
	\newcommand{\esp}{\end{split}}

\newcommand{\bdes}{\begin{description}}
	\newcommand{\edes}{\end{description}}

\newcommand{\bass}{\begin{assumption}}
	\newcommand{\eass}{\end{assumption}}
\newcommand{\bthm}{\begin{thm}}
	\newcommand{\ethm}{\end{thm}}
\newcommand{\blem}{\begin{lem}}
	\newcommand{\elem}{\end{lem}}
\newcommand{\bcor}{\begin{cor}}
	\newcommand{\ecor}{\end{cor}}
\newcommand{\bpf}{\begin{proof}}
	\newcommand{\epf}{\end{proof}}

\def\i1{_{i1}}

\def\mhf{^{-1/2}}

\def\half{^{1/2}}
\def\inv{^{-1}}

\def\diam{{\rm diam}}

\def\id{{\rm id}}

\def\diffop{\ \mathrm{d}}

\def\wh{\widehat}
\def\wt{\widetilde}
\def\wc{\widecheck}

\def\ra{\rightarrow}

\def \mbb {\mathbb}
\def \mbf {\mathbf}
\def \mc {\mathcal}

\DeclareMathOperator*{\argmin}{argmin}

\def\expect{\mbb{E}}

\def\real{\mbb{R}}
\def\ntnum{\mbb{N}}

\def\O{O}
\def\o{o}
\def\Op{O_p}

\def\gify{\rightarrow\infty}

\usepackage{ifthen}
\newcommand{\hilbert}[1][]{%
	\ifthenelse{ \equal{#1}{} }
	{\ensuremath{\mathcal{L}^2}}
	{\ensuremath{\mathcal{L}^2_{#1}}}
} 
\def\ltwoNorm#1{\|#1\|}
\newcommand{\hsSpace}[2][]{
	\ifthenelse{\equal{#1}{}}
	{\ensuremath{\mathcal H_{#2}}}
	{\ensuremath{\mathcal H_{#1,#2}}}
} 

\def\innerprod#1#2{\langle #1, #2\rangle}

\newcommand{\Metric}[3][]{%
	\ifthenelse{ \equal{#1}{} }
	{\ensuremath{d(#2,#3)}}
	{\ensuremath{d_{#1}(#2,#3)}}
}
\newcommand{\MetricSq}[3][]{%
	\ifthenelse{ \equal{#1}{} }
	{\ensuremath{d^2(#2,#3)}}
	{\ensuremath{d^2_{#1}(#2,#3)}}
}

\def\wdist#1#2{\Metric[W]{#1}{#2}}
\def\wdistSq#1#2{\MetricSq[W]{#1}{#2}}
\def\wdistnull{d_{W}}

\def\nobj{n} 
\def\objidx{i}
\def\sumn{\sum_{\objidx=1}^{\nobj}}
\newcommand{\ndp}[1][]{
	\ifthenelse{ \equal{#1}{} }
	{\ensuremath{m}}
	{\ensuremath{m_{#1}}}
}
\def\ndpi{\ndp_{\objidx}}
\def\bw{h}
\def\wsp{\mc{W}}
\def\dom{\mc D}
\def\parg{u}
\def\rarg{x}
\def\cdn{\mid}

\def\cdf{F}
\def\qt{F\inv}
\def\den{f}
\def\anyDtn{p}

\def\anyDtnEst{\wh p}
\def\cdfAnyDtn{F}
\def\qtAnyDtn{F\inv}
\def\cdfAnyDtnEst{\wh F}
\def\qtAnyDtnEst{\wh F\inv}

\def\refDtn{\anyDtn_0}

\newcommand\Exp[1][]{\mathrm{Exp}_{#1}}
\newcommand\Log[1][]{\mathrm{Log}_{#1}}

\def\jointdtn{\mc F}
\def\tdom{\mc T}
\def\tm{t}
\def\vartm{s}
\def\rtm{T}
\def\tmi{T_{\objidx}}

\def\varobjidx{\objidx'}
\def\vartmi{T_{\varobjidx}}
\def\denRtm{f_{\rtm}}

\def\dtn{P}
\def\dtni{P_{\objidx}}
\def\cdfDtniEst{\wh F_{\dtni}}
\def\dpt{X}
\def\dptidx{j}
\def\dtniEst{\wh P_{\objidx}}
\def\vardtniEst{\wh P_{\varobjidx}}
\newcommand\dtnCm[1][\tm]{\mu_\oplus(#1)} 
\def\dtnCmNull{\mu_\oplus}

\def\tansp#1{\mathscr{T}_{#1}} 

\newcommand\dtnLm[1][\tm]{\nu_\oplus(#1)} 
\newcommand\dtnLmOrc[1][\tm]{\wt\nu_\oplus(#1)} 
\newcommand\dtnLmEst[1][\tm]{\wh\nu_\oplus(#1)} 
\newcommand\dtnLmVEst[1][\tm]{\wc\nu_\oplus(#1)} 
\def\varbw{\bw'}
\def\dtnLmEstcv{\wh\nu_{\oplus \varbw}^{-\objidx}(\tmi)}

\newcommand\cdfCm[1][\tm]{F_{\dtnCm[#1]}}
\newcommand\qtCm[1][\tm]{F\inv_{\dtnCm[#1]}}
\newcommand\cdfLmEst[1][\tm]{F_{\dtnLmEst[#1]}}
\newcommand\qtLmEst[1][\tm]{F\inv_{\dtnLmEst[#1]}}
\newcommand\cdfLmVEst[1][\tm]{F_{\dtnLmVEst[#1]}}
\newcommand\qtLmVEst[1][\tm]{F\inv_{\dtnLmVEst[#1]}}

\newcommand\wtg[1][\tm]{V_{#1}}
\def\tminc{\Delta}
\def\tmincEst{\Delta}

\newcommand\wtgEst[1][\tm]{\wh V_{#1,\tmincEst}}
\newcommand\wtgVEst[1][\tm]{\wc V_{#1,\tmincEst}}

\def\bwAtom{b}
\def\ngrid{B}

\newcommand\drate[1][\ndp]{\alpha_{#1}}

\def\radUnif{\rho}
\def\tmcenter{\tm_0}
\def\ker{K}
\def\kerh{\ker_{\bw}}
\def\kmom{\kappa}
\def\momidx{z}
\def\kvar{\sigma_0^2}
\def\kmomEst{\wh\kappa}
\def\kvarEst{\wh\sigma_0^2}

\def\wLoc#1#2#3{w(#1,#2,#3)}
\def\wLocEst#1#2#3{\wh w (#1,#2,#3)}
\def\wLocDf#1{\wLoc{#1}{\tm}{\bw}} 
\def\wLocEstDf#1{\wLocEst{#1}{\tm}{\bw}}

\def\objfCm{M}
\def\objfLm{L_{\bw}}
\def\objfLmOrc{\wt L_{\nobj}}
\def\objfLmEst{\wh L_{\nobj}}

\def\ptr#1#2{\Gamma_{#1,#2}} 
\def\otr#1#2{\Upsilon_{#1,#2}} 

\def\p#1{\pi(#1)}

\def\tgaus#1#2{\mc{N}_{[0,1]}(#1,#2)}
\def\meantgaus{\zeta_{\tm}}
\def\sdtgaus{\upsilon_{\tm}}
\def\cdfStdgaus{\Phi}
\def\qtStdgaus{\Phi\inv}
\def\denStdgaus{\varphi}
\def\distort{a}
\newcommand\distortFctn[1][\distort]{g_{#1}}

\def\iid{\text{independently and identically distributed }}
\def\tcdf{cumulative distribution function\xspace}
\def\tcdfs{cumulative distribution functions\xspace}
\renewcommand{\emph}{\text}

\def\anyFctn{g}
\def\anySet{A}
\def\pushforward#1#2{#1\# #2}
\def\unif{\mathrm{Unif}}

\def\pcon{\eta} 
\def\nbw{\nobj\bw}

\def\titv{[\tmcenter-\radUnif,\tmcenter+\radUnif]}

\def\interior{^\circ}
\newcommandx*\kermom[2]{\mc{K}_{#1,#2}}
\def\const{\mathrm{const.}}

\def\wqTrue{q}
\def\wqEmp{\overline{q}}
\def\wqOrc{\wt{q}}
\def\wqEst{\wh{q}}
\def\qtdtn{\qt_{\dtn}}
\def\qtdtni{\qt_{\dtni}}
\def\qtdtniEst{\wh{F}\inv_{\dtni}}
\def\denCm{\den_{\dtnCm}}

\def\tempcdf{G}
\def\location{\xi_{\tm}}
\def\scale{\varsigma_{\tm}}

\def\references{\bibliography{merged}}

\newcommand\Author{Chen \& M\"uller}
\newcommand\Title{Wasserstein Temporal Gradients}

\title{Wasserstein Gradients for the Temporal Evolution of\\ Probability Distributions\thanks{This work was supported by NSF Grant DMS-1712864.}}
\author{Yaqing Chen \& Hans-Georg M\"uller\\
	Department of Statistics, University of California, Davis}
\date{}

\begin{document}
\maketitle
\single

\begin{abstract}
\no Many studies have been conducted on flows of probability measures, often in terms of gradient flows. We utilize a generalized notion of derivatives with respect to time to model the instantaneous evolution of empirically observed one-dimensional distributions that vary over time and develop consistent estimates for these derivatives. Employing local Fr\'echet regression and working in local tangent spaces with regard to the Wasserstein metric, we derive the rate of convergence of the proposed estimators. The resulting time dynamics are illustrated with time-varying distribution data that include yearly income distributions and the evolution of mortality over calendar years.\\  
	
\no{\textit{Key words and phrases}: Time-varying density functions, Wasserstein metric, Dynamics of income distributions, Evolution of human mortality.}
\end{abstract}

\section{Introduction}\label{sec:intro}

\begin{figure}[hbt!]
	\centering
	\includegraphics[width=.75\textwidth]{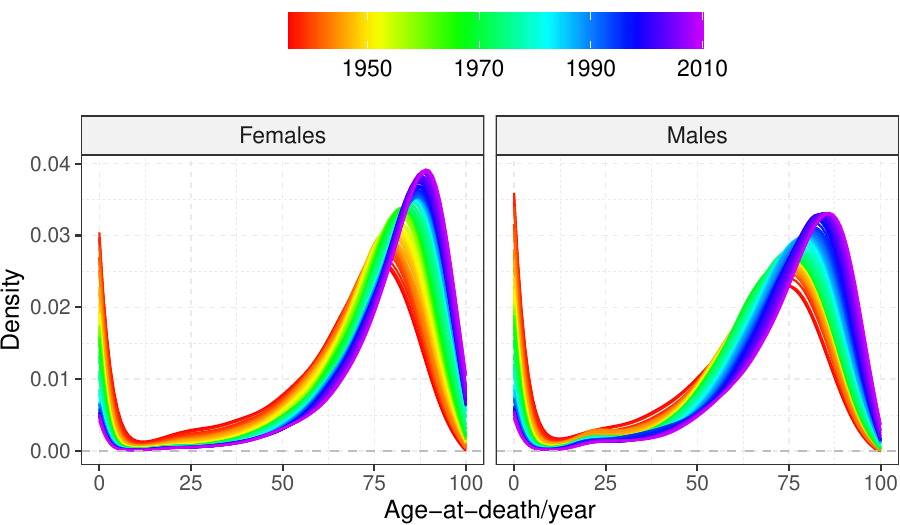}
	\caption{Time-varying densities of age-at-death (in years) for the US from \update{1936 to 2010}.}\label{fig:mort_us}
\end{figure}

There exists a  sizeable literature on flows of probability measures, often described in terms of gradient flows \citep{ambr:08,sant:17}. 
However, the statistical modeling of the instantaneous evolution of observed distributions that are indexed by time has not yet been explored. 
Figure~\ref{fig:mort_us} shows an example of time-indexed densities, which correspond to demographic age-at-death distributions from 1936 to 2010 in the US, for females and males respectively. Motivated by this and similar data, we study temporal flows for one-dimensional probability distributions. 
Recently, there has been intensive interest in comparing distributions with the Wasserstein distance, both in theory and applications \citep[e.g.][]{bols:03, bigo:17:2, gali:17, caze:18, bigo:19:2}, and in visualization   \citep[e.g.][]{deli:17}. In the one-dimensional case that we consider here, it is well known that the
Wasserstein transport can also be expressed in terms of quantile functions \citep{hoef:40,mull:11:4,chow:16}. 

Our goal is to develop statistical models that reflect instantaneous evolution of such temporal flows of distributions. 
Starting with the Monge-Kantorovich problem \citep[e.g.,][]{ambr:03:3,vill:03,vill:08}, given two probability measures $\anyDtn_1$ and $\anyDtn_2$, one aims to transport the pile of mass distributed as in $\anyDtn_1$ to that as in $\anyDtn_2$ while minimizing the transport cost. 
The transport map attaining the minimum transport cost defines the \emph{optimal transport} from $\anyDtn_1$ to $\anyDtn_2$. 
Based on such optimal transport maps, basic concepts such as tangent bundles and exponential and logarithmic maps in Riemannian manifolds can be generalized to the space of univariate probability distributions endowed with the Wasserstein distance, which form a quasi-Riemannian manifold \citep[e.g.,][]{ambr:08,bigo:17:2,zeme:19}. 
The log map, defined as the difference between the optimal transport and identity maps, captures the direction and distance of each small element of mass along the order-preserving transport from the starting probability measure to the target measure and can be used to quantify the change between the two probability measures. 
Hence, we utilize temporal derivatives of log maps, the \emph{Wasserstein temporal gradients}, to model the instantaneous temporal evolution of distributions. For this purpose, we harness local Fr\'echet regression \citep{pete:19}  to first smooth the observed probability measures over time due to the discrepancy between the true conditional Fr\'echet mean and the observed distributions  and then estimate the Wasserstein temporal gradients by difference quotients based on the local Fr\'echet regression estimates. 

The Wasserstein temporal gradients that we target  are introduced in Section~\ref{sec:method}, with estimation and asymptotic theory in Section~\ref{sec:est}. In Section \ref{sec:simu}, we discuss implementation details, followed by a simulation study. Applications are  demonstrated in Section~\ref{sec:app} for longitudinal household income and human mortality data. 

\section{Preliminaries}\label{sec:method}

\subsection{Optimal Transport in the Wasserstein Space}\label{sec:optTrans}

Given a \update{compact} interval $\dom$ in $\real$, we focus on the Wasserstein space $\wsp = \wsp(\dom)$ of distributions on $\dom$ \revtwo{(for which second moments are finite)},  endowed with the $\hilbert$-Wasserstein distance $\wdistnull$. This metric is related to the solution of Monge's optimal transport problem \citep{vill:03} and has repeatedly been rediscovered, with examples including  Mallow's distance \citep{mall:72}, earth mover's distance \citep{rubn:97} or quantile  normalization \citep{bols:03}.

Specifically, the $\hilbert$-Wasserstein distance between any $\anyDtn_1,\anyDtn_2\in\wsp$ is the square root of 
\be \label{eq:monge} \wdistSq{\anyDtn_1}{\anyDtn_2} = \inf_{\pushforward{\anyFctn}{\anyDtn_1} = \anyDtn_2} \int_{\dom} [\rarg - \anyFctn(\rarg)]^2 \diffop\anyDtn_1(\rarg), \ee
where $\pushforward{\anyFctn}{\anyDtn}$ is a push-forward measure such that $\pushforward{\anyFctn}{\anyDtn}(\anySet) = \anyDtn(\{\rarg: \anyFctn(\rarg)\in\anySet\})$, for any measurable function $\anyFctn\colon \real\ra\real$, distribution $\anyDtn\in\wsp$ and set $\anySet\subseteq\real$. 
It is well known \citep[e.g.,][]{camb:76} that \update{if $\anyDtn_1$ is atomless, the minimum in \eqref{eq:monge} is  attained at the optimal transport map $\otr{\anyDtn_1}{\anyDtn_2} = \qtAnyDtn_2\circ\cdfAnyDtn_1$ from $\anyDtn_1$ to $\anyDtn_2$, and is 
	$\wdistSq{\anyDtn_1}{\anyDtn_2} = \int_0^1 [\qt_1(\parg) - \qt_2(\parg)]^2\diffop \parg$. 
	Here, $\cdf_l$ and $\qt_l$ are the cumulative distribution function 
	and quantile function of $\anyDtn_l$ for $l=1,2$, where \tcdfs are considered to be  right continuous and quantile functions to be left continuous. 
	\revtwo{A distribution $\anyDtn$ is atomless if it has  a continuous cumulative distribution function.}
	
	Basic concepts of Riemannian manifolds can be analogously defined in the Wasserstein space $\wsp$ based on optimal transport maps \citep[e.g.,][]{ambr:08,bigo:17:2,zeme:19}. 
	\revtwo{Suppose that $\refDtn$ is an atomless reference probability measure in $\wsp$.}  
	The tangent space at $\refDtn$ is defined as 
	\citep[Equation (8.5.1),][]{ambr:08}
	\[ \tansp{\refDtn} = {\overline{\{\pcon(\otr{\refDtn}{\anyDtn} - \id): \anyDtn\in\wsp,\ \pcon>0\}}}^{\hilbert[\refDtn]}, \]
	where $\hilbert_{\refDtn} = \hilbert_{\refDtn}(\dom)$ is the Hilbert space of $\refDtn$-square-integrable functions on $\dom \subset \real$, with inner product $\innerprod{\cdot}{\cdot}_{\refDtn}$ and norm $\ltwoNorm{\cdot}_{\refDtn}$; \update{we reserve the notations without subscripts $\innerprod{\cdot}{\cdot}$ and $\ltwoNorm{\cdot}$ for the the inner product and norm corresponding to the Lebesgue measure}. 
	Due to the atomlessness of $\refDtn$, the tangent space $\tansp{\refDtn}$ is a subspace of $\hilbert[\refDtn]$ equipped with the same inner product and induced norm. 
	The exponential map $\Exp[\refDtn]: \tansp{\refDtn}\ra \wsp$ is then defined by 
	\[ \Exp[\refDtn]\anyFctn = \pushforward{(\anyFctn+\id)}{\refDtn}, \quad\text{for }\anyFctn \in\tansp{\refDtn}. \]
	Although the exponential map here is not a local homeomorphism as in Riemannian manifolds \citep{ambr:04}, 
	any $\anyDtn\in\wsp$ can be recovered from $\refDtn$ by $\Exp[\refDtn](\otr{\refDtn}{\anyDtn}-\id)$, 
	which motivates the definition of the inverse of the exponential map, i.e., the logarithmic map $\Log[\refDtn]: \wsp\ra \tansp{\refDtn}$,
	\[\Log[\refDtn]\anyDtn = \otr{\refDtn}{\anyDtn} -\id, \quad\text{for }\anyDtn\in\wsp. \]
	The tangent vector given by log maps quantifies the difference between $\refDtn$ and $\anyDtn$. Indeed, $\ltwoNorm{\Log[\refDtn]\anyDtn}_{\refDtn} = \wdist{\refDtn}{\anyDtn}$. 
	Furthermore, the difference between optimal transport maps and the identity map reveals how mass is transported between distributions provided that the order is preserved. 
	Specifically, given $\rarg\in\dom$, if $\otr{\anyDtn_1}{\anyDtn_2}(\rarg) > \rarg$ (respectively, $\otr{\anyDtn_1}{\anyDtn_2}(\rarg) < \rarg$), then $\rarg$ should be moved to the right (respectively, left) to $\otr{\anyDtn_1}{\anyDtn_2}(\rarg)$ in order to keep its rank, i.e., $\cdfAnyDtn_2(\otr{\anyDtn_1}{\anyDtn_2}(\rarg)) = \cdfAnyDtn_1(\rarg)$. 
	
	\subsection{Wasserstein Temporal Gradients}\label{sec:wtg}
	
	Let $(\rtm,\dtn)$ be a pair of random elements in $\tdom\times \wsp$ with joint distribution $\jointdtn$, where $\tdom\subseteq\real$ is the time domain. We assume
	\ben[label=(A\arabic*), series = dtn]
	\item\label{ass:atomless} \update{$\dtn$ is atomless almost surely.}
	\een
	Note that $\expect[\wdistSq{\dtn}{\anyDtn} \cdn \rtm = \tm] \le \diam(\dom)^2<\infty$ for all $\anyDtn\in\wsp$ and $\tm\in\tdom$. 
	Since the Wasserstein space $\wsp$ is a Hadamard space \citep{kloe:10}, there exists a unique minimizer of $\expect[\wdistSq{\dtn}{\cdot} \cdn \rtm = \tm]$ \citep{stur:03}. 
	Thus, the conditional Fr\'echet mean $\dtnCm$ of $\dtn$ given $\rtm = \tm$ is well-defined; specifically, 
	\bal\nn 
	\dtnCm = \argmin_{\anyDtn\in\wsp} \objfCm(\anyDtn,\tm), \quad\text{with } \objfCm(\anyDtn,\tm) \coloneqq \expect[\wdistSq{\dtn}{\anyDtn}\cdn\rtm = \tm], \eal
	\update{and the quantile function of $\dtnCm$ is given by $\qtCm(\cdot) = \expect[\qtdtn(\cdot) \cdn\rtm=\tm]$, where $\qtdtn$ is the quantile function of $\dtn$.} 
	
	To model the instantaneous temporal evolution of probability distributions, we are aiming to generalize the notion of derivatives, which are used to quantify the instantaneous change of differentiable real-valued functions, to the scenario of temporal distribution flows. 
	As discussed in Section~\ref{sec:optTrans}, log maps quantify the discrepancy between two probability distributions. 
	\update{We note that the atomlessness of $\dtnCm$ is guaranteed by \ref{ass:atomless}.} 
	Hence, a measure of instantaneous temporal evolution of distributions, the \emph{Wasserstein temporal gradient} at time $\tm\in\tdom$ can be defined by 
	\bal\label{eq:wtg} 
	\wtg &= \lim_{\tminc\ra 0}\frac{\Log[\dtnCm]\dtnCm[\tm+\tminc]}{\tminc} \\
	&= \lim_{\tminc\ra 0} \frac{\qtCm[\tm+\tminc]\circ\cdfCm - \id}{\tminc}\\ 
	&= \frac{\partial \qtCm}{\partial\tm}\circ \cdfCm, \quad \revtwo{\dtnCm\text{-a.e.}}, \eal
	provided that \revtwo{the bivariate function $(\tm,\parg)\mapsto\qtCm(\parg)$ is differentiable with respect to $\tm$}. 
	Here, $\cdfCm[\vartm]$ and $\qtCm[\vartm]$ are the \tcdf and quantile function of $\dtnCm[\vartm]$ for $\vartm\in\tdom$. 
	If there exists $g\in\mc L^1(\tdom)$ such that  $\wdist{\dtnCm[\vartm]}{\dtnCm} \le \int_{\vartm}^{\tm} g(\rarg)\diffop\rarg$, then $\dtnCmNull$ is an absolutely continuous curve  in the Wasserstein space, and $\wtg$ is also referred to as the velocity vector of $\dtnCmNull$ \citep{ambr:04}. 
	
	\begin{eg}\label{eg:wtg_eg2}
		For $\tm\in\tdom$, let $\dtnCm = \tgaus{\meantgaus}{\sdtgaus^2}$ be a truncated Gaussian distribution on the interval $[0,1]$. Then the Wasserstein temporal gradient at $\tm$ is
		\bal\nn
		\wtg(\rarg) &= \meantgaus' + (\rarg-\meantgaus) \frac{\sdtgaus'}{\sdtgaus} \\
		&\quad - \sdtgaus \frac{\cdfCm(\rarg) \left[\frac{\sdtgaus'}{\sdtgaus} +  \left(\frac{\meantgaus}{\sdtgaus}\right)'\right] \denStdgaus\left(\frac{1-\meantgaus}{\sdtgaus}\right) +(1-\cdfCm(\rarg) \left(\frac{\meantgaus}{\sdtgaus}\right)' \denStdgaus\left(\frac{-\meantgaus}{\sdtgaus}\right)}{\denStdgaus \circ \qtStdgaus \left( \cdfCm(\rarg) \cdfStdgaus\left(\frac{1-\meantgaus}{\sdtgaus}\right) + (1-\cdfCm(\rarg)) \cdfStdgaus\left(\frac{-\meantgaus}{\sdtgaus}\right)\right)}, \eal
		where $\cdfCm(\rarg) = [\cdfStdgaus((\rarg-\meantgaus)/\sdtgaus) - \cdfStdgaus(-\meantgaus/\sdtgaus)] / [\cdfStdgaus((1-\meantgaus)/\sdtgaus) - \cdfStdgaus(-\meantgaus/\sdtgaus)]$, $\cdfStdgaus$ and $\denStdgaus$ are the \tcdf and density of standard normal distributions, respectively, and we use the notation $\anyFctn_{\tm}'= (\mathrm{d}/\mathrm{d}\tm)\anyFctn_{\tm}=(\mathrm{d}/\mathrm{d}\tm) \anyFctn(\tm)$ for a function $\anyFctn$. \update{Densities and Wasserstein temporal gradients of $\tgaus{\meantgaus}{\sdtgaus^2}$ with different values of $\meantgaus$ and $\sdtgaus$ are shown in Figure~\ref{fig:eg_tgaus}}.
	\end{eg}
	
	\begin{figure}[hbt!]
		\centering
		\begin{subfigure}[c]{0.65\textwidth}
			\includegraphics[width=\linewidth]{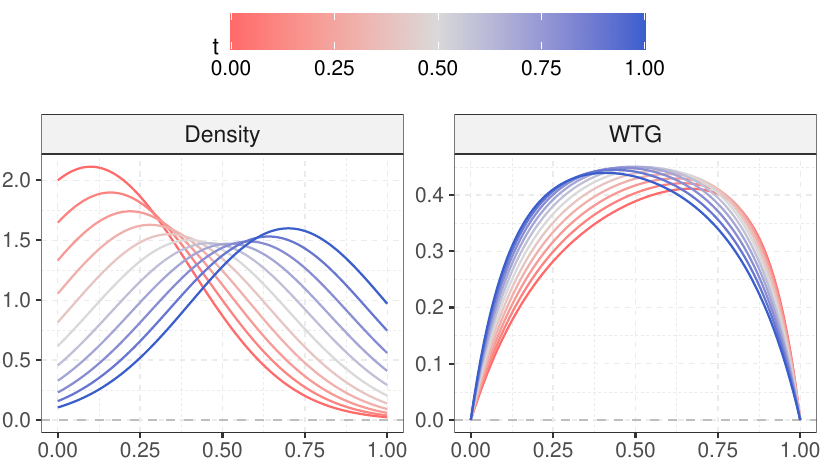}
			\caption{$\meantgaus = 0.1 + 0.6\tm$, $\sdtgaus \equiv 0.3$.}
		\end{subfigure}\\\vspace{1em}
		\begin{subfigure}[c]{0.65\textwidth}
			\includegraphics[width=\linewidth]{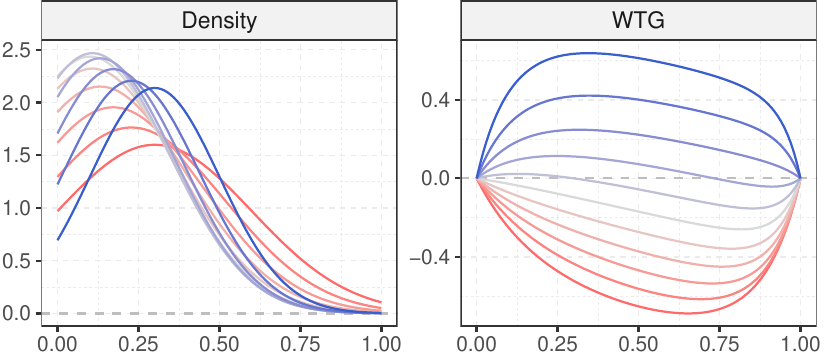}
			\caption{$\meantgaus = 0.1 + 0.8 (\tm-0.5)^2$, $\sdtgaus = 0.3 - 0.1 \tm$.}
		\end{subfigure}
		\caption{Densities and Wasserstein temporal gradients of $\dtnCm = \tgaus{\meantgaus}{\sdtgaus^2}$, for $\tm\in[0,1]$.}\label{fig:eg_tgaus}
	\end{figure}
	
	\update{
		\begin{eg}
			For $\tm\in\tdom$, let $\dtnCm$ be atomless distributions in a location-scale family with location and scale parameters being $\location$ and $\scale$, respectively. 
			Specifically, the \tcdf of $\dtnCm$ is given by $\rarg\mapsto \tempcdf((\rarg-\location)/\scale)$, where $\tempcdf\colon\real\ra [0,1]$ is a template \tcdf. 
			Then the Wasserstein temporal gradient at $\tm\in\tdom$ is 
			\bal\nn
			\wtg(\rarg) = \location' + (\rarg-\location) \frac{\scale'}{\scale}. \eal 
		\end{eg}
	}
	
	For a real-valued differentiable function $\anyFctn:\tdom\ra\real$, 
	\[\frac{\diffop \cdfCm(\anyFctn(\tm))}{\diffop \tm} = 0\quad \text{if and only if}\quad \anyFctn'(\tm) = \wtg(\anyFctn(\tm)).\]
	Thus, comparing the actual flow $\anyFctn'(\tm)$ for a given longitudinal trajectory $\anyFctn(\tm)$ with the optimal flow $\wtg(\anyFctn(\tm))$ provides insights into how the rank of $\anyFctn(\tm)$ changes at each time $\tm$. 
	If $\anyFctn'(\tm)>\wtg(\anyFctn(\tm))$ (respectively, $\anyFctn'(\tm)<\wtg(\anyFctn(\tm))$), then $\frac{\diffop}{\diffop \tm } \cdfCm(\anyFctn(\tm))$ is positive (respectively, negative), i.e., the rank of $\anyFctn(\tm)$ increases (respectively, decreases) instantaneously at time $\tm$.
	
	\bco
	\begin{rem}
		Compared with an alternative, the pointwise derivatives of densities with respect to time, which we refer to as density temporal gradients (DTGs), the Wasserstein temporal gradient (WTG) reveals different features of the evolution of distribution flows, as it captures the direction and speed of the mass flow in a neighborhood of $\rarg\in\dom$. In contrast, the DTG characterizes how fast the density instantaneously tends to increase or decrease for any $\rarg$ in the support of the current distribution. The interpretation of the WTG is more intuitive in terms of transportation of mass and it is better suited to reflect the salient features of density evolution. 
	\end{rem}
	\fi
	
	\section{Estimation and Theory}\label{sec:est}
	\subsection{Distribution Estimation}\label{sec:dtnEst}
	In practice, distributions are usually not fully observed. This creates an additional challenge for the implementation of the Wasserstein temporal gradients. 
	This issue can be addressed, for example, by estimating \tcdfs  \citep[e.g.,][]{agga:55,read:72,falk:83,lebl:12}, or estimating quantile functions \citep[e.g.,][]{parz:79,falk:84,yang:85,chen:97:2} of the underlying distributions from which the observed data are sampled. 
	Note that with any quantile function estimator $\qtAnyDtnEst$ (respectively, \tcdf estimator $\cdfAnyDtnEst$), the corresponding \tcdf (respectively, quantile function) can be obtained by right (respectively, left) continuous inversion, 
	\begin{gather*}
	\cdfAnyDtnEst(\rarg) = \sup\{\parg\in[0,1]:\qtAnyDtnEst(\parg)\le \rarg\},\quad\text{for }\rarg\in\real, \\
	\quad\quad\quad\quad \text{     respectively, } \qtAnyDtnEst (\parg) = \inf\{\rarg\in\dom: \cdfAnyDtnEst(\rarg)\ge \parg\},\quad\text{for }\parg\in(0,1). 
	\end{gather*}
	Alternatively, one can first estimate densities \citep{pana:16,pete:16:1} and then obtain the \tcdfs and quantile functions by integration and inversion. 
	
	Suppose $\{(\tmi,\dtni)\}_{\objidx=1}^{\nobj}$ are $\nobj$ independent realizations of $(\rtm,\dtn)$. 
	Available observations are samples of independent measurements $\{\dpt_{\objidx\dptidx}\}_{\dptidx=1}^{\ndpi}$ generated from $\dtni$, respectively, 
	where $\ndpi$ are the sample sizes which may vary across distributions $\dtni$, for $\objidx = 1,\ldots,\nobj$. 
	Note that the observed data $\dpt_{\objidx\dptidx}$ result from two independent random mechanisms:  The first of these generates independently and identically distributed pairs $(\tmi,\dtni)$; the second generates samples of observations $\{\dpt_{\objidx\dptidx}\}_{\dptidx = 1}^{\ndpi}$ according to each distribution $\dtni$, i.e., $\dpt_{\objidx\dptidx}\sim \dtni$ independently. 
	
	For a given distribution $\anyDtn\in\wsp$, with a \tcdf estimate $\cdfAnyDtnEst$ obtained by any estimation method based on a random sample generated from $\anyDtn$, we denote by $\anyDtnEst = \p{\cdfAnyDtnEst}$ the distribution associated with $\cdfAnyDtnEst$. 
	We make the following assumption on the discrepancy of the estimated and true probability distributions for the theoretical analysis of the Wasserstein temporal gradient estimation.    
	\ben[label = (D\arabic*), series = dtnEst]
	\item \label{ass:distn_est_rate}
	For any distribution $\anyDtn\in\wsp$, with nonnegative decreasing sequence $\drate = o(1)$  as $\ndp\ra\infty$, the corresponding estimate $\anyDtnEst$ based on a sample of size $\ndp$ generated from $\anyDtn$ satisfies
	\bgt\nn 
	\sup_{\anyDtn\in\wsp} \expect[\wdistSq{\anyDtnEst}{\anyDtn}]= O(\drate).\egt
	\een
	We note that this assumption can be easily satisfied. For example, the density estimator proposed by \citet{pana:16} satisfies \ref{ass:distn_est_rate} with $\drate = \ndp\mhf$. 
	If only considering the distributions in $\wsp$ with densities satisfying 
	\bgt\nn \sup_{\rarg\in\revtwo{\mathrm{supp}(\den_{\anyDtn})}} \max\{\den_{\anyDtn}(\rarg), 1/\den_{\anyDtn}(\rarg), |\den_{\anyDtn}'(\rarg)|\} \le C,\text{ uniform across }\anyDtn, \egt
	where $\den_{\anyDtn}$ is the density function of a distribution $\anyDtn\in\wsp$, \revtwo{$\mathrm{supp}(\den_{\anyDtn})=\dom$ is the support of $\anyDtn$ and $C>0$ is a constant, 
		then the empirical measure satisfies \ref{ass:distn_est_rate} with $\drate = \ndp\inv$}. 
	
	In order to deal with the estimation of $\nobj$ distributions simultaneously, we also require
	\ben[label = (D\arabic*), resume=dtnEst]
	\item \label{ass:nObsPerDtn}
	There exists a sequence $\ndp = \ndp(\nobj)$ such that $\min_{1\le \objidx \le\nobj} \{\ndpi\}\ge \ndp$ and $\ndp\ra\infty$ as $\nobj\ra \infty$.
	\een
	
	\subsection{Estimation of Wasserstein Temporal Gradients}\label{sec:wtgEst}
	We assume that for each $\objidx = 1,\ldots,\nobj$, we obtain an estimate $\cdfDtniEst$ of the \tcdf of $\dtni$ by one of the methods discussed in Section~\ref{sec:dtnEst} from the observed data $\{\dpt_{\objidx\dptidx}\}_{\dptidx=1}^{\ndpi}$. 
	Denote by $\dtniEst = \p{\cdfDtniEst}$ the distribution associated with $\cdfDtniEst$. 
	Since the discrepancy $\expect[\wdistSq{\dtni}{\dtnCm[\tmi]}\cdn \tmi]$ between the random distributions $\dtni$ and the conditional Fr\'echet means $\dtnCm[\tmi]$ does not vanish as $\nobj\gify$, difference quotients based on the estimated distributions $\dtniEst$ are not directly suitable as an estimate of Wasserstein temporal gradients. 
	
	Accordingly,  we utilize local Fr\'echet regression \citep{pete:19} to smooth the distributions $\{\dtniEst\}$ over time, which yields consistent estimates of $\dtnCm$, for any $\tm\in\tdom$. 
	Following \citet{pete:19}, we define the localized Fr\'echet mean by
	\be \label{eq:dtnLm}
	\dtnLm = \argmin_{\anyDtn\in\wsp}\objfLm(\anyDtn,\tm),\quad\text{with } \objfLm(\anyDtn,\tm) = \expect[\wLocDf{\rtm} \wdistSq{\dtn}{\anyDtn}]. \ee 
	Here, $\wLocDf{\vartm} = \kerh(\vartm-\tm)[\kmom_2(\tm) - \kmom_1(\tm)(\vartm - \tm)]/\kvar(\tm)$,
	where $\kmom_{\momidx}(\tm) = \expect[\kerh(\rtm - \tm)(\rtm - \tm)^{\momidx}]$, for $\momidx=0,1,2$, $\kvar(\tm) = \kmom_0(\tm)\kmom_2(\tm) - \kmom_1(\tm)^2$, $\kerh(\cdot) = \ker(\cdot/\bw)/\bw$, $\ker$ is a smoothing kernel, i.e., a density function symmetric around zero, and $\bw = \bw(\nobj)>0$ is a bandwidth sequence. 
	If assuming the distributions $\dtni$ are fully observed, setting 
	$\wLocEstDf{\vartm} = \kerh(\vartm - \tm)[\kmomEst_2(\tm) - \kmomEst_1(\tm)(\vartm - \tm)]/\kvarEst(\tm)$, 
	where $\kmomEst_{\momidx}(\tm) = {\nobj}\inv\sumn\kerh(\tmi - \tm)(\tmi - \tm)^{\momidx}$, for $\momidx = 0,1,2$, and $\kvarEst(\tm) = \kmomEst_0(\tm)\kmomEst_2(\tm) - \kmomEst_1(\tm)^2$, 
	an oracle local Fr\'echet regression estimate is
	\be \label{eq:dtnLmOrc}
	\dtnLmOrc =  \argmin_{\anyDtn\in\wsp}\objfLmOrc(\anyDtn,\tm),\quad\text{with } \objfLmOrc(\anyDtn,\tm) = {\nobj}\inv \sumn \wLocEstDf{\tmi} \wdistSq{\dtni}{\anyDtn}. \ee 
	In practice, we usually only observe random samples of measurements $\dpt_{\objidx\dptidx}$ generated from $\dtni$. 
	Replacing $\dtni$ with the corresponding estimates $\dtniEst$ as discussed in Section \ref{sec:dtnEst}, a data-based local Fr\'echet regression estimate is 
	\be\label{eq:dtnLmEst}
	\dtnLmEst = \argmin_{\anyDtn\in\wsp}\objfLmEst(\anyDtn,\tm),\quad\text{with } \objfLmEst(\anyDtn,\tm) = {\nobj}\inv\sumn \wLocEstDf{\tmi} \wdistSq{\dtniEst}{\anyDtn}. \ee 
	
	For simplicity, we assume for theoretical analysis that the support of the marginal density $\denRtm$ of $\rtm$, i.e., $\tdom = \{\tm\in\real: \denRtm(\tm)>0\}$, is connected. Let $\tdom\interior$ be the interior of $\tdom$. 
	Furthermore, we require the following assumptions \update{for the asymptotic analysis of the weights $\wLocDf{\rtm}$ and $\wLocEstDf{\tmi}$ in $\dtnLm$ and $\dtnLmEst$, respectively}. 
	
	\ben[label = (R\arabic*)]
	\item \label{ass:ker} 
	\update{The kernel $\ker$ is a probability density function, symmetric around zero and continuous on $[-1,1]$, such that  $\ker(\rarg)=0$, for all $|\rarg|>1$}. 
	\item \label{ass:jointdtn} 
	The marginal density $\denRtm$ of $\rtm$ exists and is continuous on $\tdom$ and twice continuously differentiable on $\tdom\interior$. 
	The second-order derivative $\denRtm''$ is bounded,  $\sup_{\tm\in\tdom\interior}|\denRtm''(\tm)|<\infty$. 
	\een
	
	For any $\tm\in\tdom\interior$, with the local Fr\'echet regression estimate $\dtnLmEst$ as per \eqref{eq:dtnLmEst} and some small $\tmincEst>0$, an estimate of the Wasserstein temporal gradient $\wtg$ in \eqref{eq:wtg} is then given by 
	\bgt\label{eq:wtgEst} 
	\wtgEst = \frac{\qtLmEst[\tm+\tmincEst]  \circ\cdfLmEst - \id}{\tmincEst}, \egt
	where $\cdfLmEst[\vartm]$ and $\qtLmEst[\vartm]$ are the \tcdf and quantile function of $\dtnLmEst[\vartm]$ for $\vartm\in\tdom$. 
	
	\subsection{Parallel Transport}\label{sec:paraTrans}
	
	Note that the true and estimated Wasserstein temporal gradients lie in different tangent spaces; specifically, $\wtg\in \tansp{\dtnCm}$ and $\wtgEst\in\tansp{\dtnLmEst}$. To quantify the estimation discrepancy of $\wtgEst$, an expedient tool is parallel transport, which is commonly used for manifold-valued data \citep[e.g.,][]{yuan:12,lin:18,pete:19:2,chen:19:4}. 
	For two probability measures $\anyDtn_1,\anyDtn_2\in\wsp$, a parallel transport operator $\ptr{\anyDtn_1}{\anyDtn_2}\colon \hilbert[\anyDtn_1]\ra\hilbert[\anyDtn_2]$ is defined by 
	\be\nn
	\ptr{\anyDtn_1}{\anyDtn_2} \anyFctn = \anyFctn\circ \qtAnyDtn_1\circ\cdfAnyDtn_2,\quad \text{for }\anyFctn\in\hilbert[\anyDtn_1], \ee 
	where $\cdfAnyDtn_2$ and $\qtAnyDtn_1$ are the \tcdf of $\anyDtn_2$ and quantile function of $\anyDtn_1$, respectively.  
	
	Note that since $\anyDtn_1$ and $\anyDtn_2$ are atomless, the tangent spaces satisfy  $\tansp{\anyDtn_k}\subset \hilbert[\anyDtn_k]$ for $k=1,2$, and  the parallel transport operator $\ptr{\anyDtn_1}{\anyDtn_2}|_{\tansp{\anyDtn_1}}$  restricted to the tangent space $\tansp{\anyDtn_1}$  defines the parallel transport between tangent spaces $\tansp{\anyDtn_1}$ and $\tansp{\anyDtn_2}$.  
	Furthermore, the parallel transport operator $\ptr{\anyDtn_2}{\anyDtn_1}$ from $\hilbert[\anyDtn_2]$ to $\hilbert[\anyDtn_1]$ is the adjoint operator of $\ptr{\anyDtn_1}{\anyDtn_2}$, i.e., 
	$\innerprod{\ptr{\anyDtn_1}{\anyDtn_2}\anyFctn_1}{\anyFctn_2}_{\anyDtn_2} = \innerprod{\anyFctn_1}{\ptr{\anyDtn_2}{\anyDtn_1}\anyFctn_2}_{\anyDtn_1}$. 
	Thus, the discrepancy between functions $\anyFctn_1\in\tansp{\anyDtn_1}$ and $\anyFctn_2\in\tansp{\anyDtn_2}$ can be quantified by $\ltwoNorm{\ptr{\anyDtn_2}{\anyDtn_1}\anyFctn_2 - \anyFctn_1}_{\anyDtn_1}$. 
	
	\subsection{Asymptotic Theory}\label{sec:theory}
	
	As discussed in Section~\ref{sec:paraTrans}, in order to justify $\ltwoNorm{\ptr{\dtnLmEst}{\dtnCm} \wtgEst - \wtg}_{\dtnCm}$ as a measure of estimation discrepancy of $\wtgEst$, \update{we require the atomlessness of $\dtnCm$ and $\dtnLmEst$. The former follows from \ref{ass:atomless}}. 
	However, the latter is not guaranteed in general. 
	For theoretical derivations, we instead consider an atomless variant $\dtnLmVEst$ of $\dtnLmEst$, which is  defined as follows. 
	Suppose $\min\dom = \rarg_0 < \rarg_1 <\dots < \rarg_{\ngrid} = \max\dom$ is an equidistant grid on $\dom$ with increment $\bwAtom$. Then the \tcdf of $\dtnLmVEst$ is given by $\cdfLmVEst(\rarg) = \cdfLmEst(\rarg_{l-1}) + \bwAtom\inv (\rarg - \rarg_{l-1})[\cdfLmEst(\rarg_l) - \cdfLmEst(\rarg_{l-1})]$, for $\rarg\in[\rarg_{l-1},\rarg_l)$; $\cdfLmVEst(\rarg) =0$ and 1 for $\rarg < \rarg_0$ and $\rarg\ge\rarg_{\ngrid}$, respectively. 
	We assume that $\bwAtom = \bwAtom(\nobj)$ is a positive sequence such that $\bwAtom\ra 0$ as $\nobj\gify$. 
	Hence, an estimate of the Wasserstein temporal gradient at time $\tm$ based on $\dtnLmVEst[\vartm]$ with $\vartm\in\tdom$ is given by
	\bgt\nn 
	\wtgVEst = \frac{\qtLmVEst[\tm+\tmincEst]\circ \cdfLmVEst - \id}{\tmincEst}. \egt
	
	To obtain the convergence rate of $\wtgVEst$, we also require the following assumption.  
	\ben[label = (A\arabic*), resume = dtn]
	\item \label{ass:dtnCm}
	\revtwo{
		The bivariate function $(\tm,\parg)\mapsto\qtCm(\parg)$ is twice differentiable and $(\tm,\parg)\mapsto\partial^2\qtCm(\parg)/(\partial\tm\partial\parg)$ is continuous with respect to $\parg$. 
		There exists a constant $C>0$ such that $\sup_{\rarg\in\dom}\denCm(\rarg) \le C$, $\int_0^1 \sup_{\tm\in\tdom}|\partial^2\qtCm(\parg)/\partial\tm^2|^2 \diffop\parg\le C$, 
		and $\sup_{\tm\in\tdom,\,\parg\in(0,1)}|\partial^2\qtCm(\parg)/(\partial\tm\partial\parg)|\le C$.}
	\een

	\update{
		We take $\tmincEst=\bw$; this choice, together with suitable values for $\bw$, $\bwAtom$ and $\tmincEst$, will lead to   Wasserstein temporal gradient estimates $\wtgVEst$ with an asymptotic rate of convergence that matches the well-known optimal rate of derivative estimation for nonparametric regression for the case of real-valued responses \revtwo{assuming twice continuous differentiability of the regression function}. This optimal rate is for example achieved by derivative estimates based on local polynomial fitting \citep{mull:87:4,fan:96}. 
		\bthm\label{thm:wtgEst}
		Assume \ref{ass:atomless}--\ref{ass:dtnCm}, \ref{ass:distn_est_rate}--\ref{ass:nObsPerDtn} and \ref{ass:ker}--\ref{ass:jointdtn}. With
		$\tmincEst=\bw$, if $\bw\ra 0$, $\nobj\bw^3\ra\infty$, $\bwAtom\bw\inv\ra\infty$, and $\drate\bw\inv\ra 0$, 
		\bal\nn
		&\left\ltwoNorm{\ptr{\dtnLmVEst}{\dtnCm}\wtgVEst - \wtg\right}_{\dtnCm}\\
		&\quad= \left\ltwoNorm{\frac{\qtLmVEst[\tm+\bw] - \qtLmVEst}{\bw} - \frac{\partial\qtCm}{\partial\tm}\right}\\
		&\quad= \O\left(\revtwo{\bw}\right) + \Op\left((\nobj\bw^3)\mhf\right) + \O\left(\bwAtom\bw\inv\right) + \Op\left((\drate\bw\inv)\half\right).\eal
		Furthermore, with $\bw\sim\nobj^{-1/\revtwo{5}}$, $\bwAtom=\O(\nobj^{-\revtwo{2/5}})$ and $\drate=\O(\nobj^{-\revtwo{3/5}})$, 
		\bal\label{eq:wtg_rate} \left\ltwoNorm{\ptr{\dtnLmVEst}{\dtnCm}\wtgVEst - \wtg\right}_{\dtnCm} = \Op\left(\nobj^{-\revtwo{1/5}}\right).\eal 
		\ethm}

	Proofs are in the appendix. 

	\section{Implementation and Simulations}\label{sec:simu}
	
	There are two tuning parameters for implementation of Wasserstein temporal gradients, namely the bandwidth $\bw$ involved in the local Fr\'echet regression as per \eqref{eq:dtnLmEst} and the time increment $\tmincEst$ used in the difference quotient estimator as per \eqref{eq:wtgEst}. 
	\update{As suggested by the theoretical analysis in Section~\ref{sec:theory}, we take $\tmincEst=\bw$ in practice. }
	We choose the bandwidth $\bw$ by leave-one-out cross validation, where the objective function to be minimized is the mean discrepancy between the local Fr\'echet regression estimates and the observed distributions; specifically, 
	\[\bw = \argmin_{\varbw} \nobj\inv\sumn \wdistSq{\dtnLmEstcv}{\dtniEst}, \]
	where $\dtnLmEstcv$ is the local Fr\'echet regression estimate of $\dtnCm[\tmi]$ obtained with bandwidth $\varbw$ based on the sample excluding the $\objidx$th pair $(\tmi,\dtniEst)$, i.e., 
	\[\dtnLmEstcv = \argmin_{\anyDtn\in\wsp} \frac{1}{\nobj-1} \sum_{\varobjidx\ne\objidx} \wLocEst{\vartmi}{\tmi}{\varbw} \wdistSq{\vardtniEst}{\anyDtn},\]
	and $\dtniEst$ is the estimate of $\dtni$ based on the observed measurements $\{\dpt_{\objidx\dptidx}\}_{\dptidx= 1}^{\ndpi}$ as discussed in Section \ref{sec:dtnEst}. 
	In practice, we replace leave-one-out cross validation by 10-fold cross validation when $\nobj > 30$. 
	
	We generated data for simulations as follows: 
	\ben[Step 1:, itemindent=1em]
	\item Set $\dtnCm = \tgaus{\meantgaus}{\sdtgaus^2}$, a truncated Gaussian distribution on $[0,1]$ with  \update{$\meantgaus = (t-0.2)(t-0.5)(t-0.9)+0.2$ and $\sdtgaus = 0.15 + 0.03 \sin(2\pi t)$}. 
	\item Sample $\distort_{\objidx}\sim \unif\{\pm 10\pi,\pm 11\pi, \ldots, \pm 14\pi\}$ and $\tmi \sim \unif[0,1]$ independently, for $\objidx = 1,\ldots,\nobj$. 
	Set $\dtni =\pushforward{\distortFctn[\distort_{\objidx}]}{\dtnCm[\tmi]}$, 
	where $\distortFctn(\rarg) = \rarg - |\distort|\inv\sin(\distort\rarg)$ with $\distort\in\real\backslash\{0\}$ and $\rarg\in\real$. 
	\item Draw an \iid sample $\{\dpt_{\objidx\dptidx}\}_{\dptidx=1}^{\ndp}$ of size $\ndp$ from each of the distributions $\{\dtni\}_{\objidx=1}^{\nobj}$.
	\een
	
	\update{Four cases were considered with $\nobj\in\{50, 200\}$ and $\ndp\in \{25, 500\}$. We simulated 500 runs for each pair $(\nobj, \ndp)$. 
		To evaluate the performance of the Wasserstein temporal gradient estimate based on the local Fr\'echet regression as per \eqref{eq:wtgEst}, we computed the integrated error (IE) for given $\tm\in[0,1]$; specifically,
		\bgt\label{eq:simu_ad} 
		\mathrm{IE}(\nobj,\ndp,\tm) = \left\ltwoNorm{\frac{\qtLmEst[\tm+\tmincEst] - \qtLmEst}{\tmincEst} - \frac{\partial}{\partial\tm}\qtCm\right}. \egt
		The results are summarized in the boxplots of IEs in Figure~\ref{fig:simu_ad}.} 
	It can be seen that the estimation error decreases as $\nobj$ or $\ndp$ increases. 
	
	\begin{figure}[hbt!]
		\centering
		\begin{subfigure}[c]{.24\textwidth}
			\includegraphics[width=.95\linewidth]{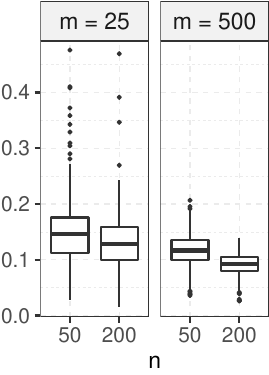}
			\caption{$\tm=0.25$.}\label{fig:simu_ad1}
		\end{subfigure}
		\begin{subfigure}[c]{.24\textwidth}
			\includegraphics[width=.95\linewidth]{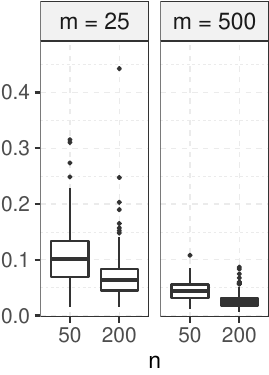}
			\caption{$\tm=0.5$.}\label{fig:simu_ad2}
		\end{subfigure}
		\begin{subfigure}[c]{.24\textwidth}
			\includegraphics[width=.95\linewidth]{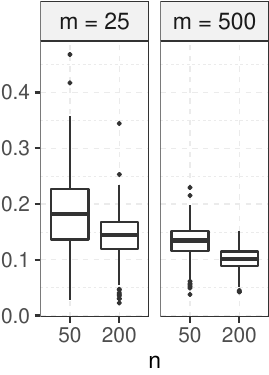}
			\caption{$\tm=0.75$.}\label{fig:simu_ad3}
		\end{subfigure}
		\caption{Boxplots of the integrated errors (IEs) as per \eqref{eq:simu_ad} of the 500 runs for $\tm\in\{0.25, 0.5, 0.75\}$ and each $(\nobj,\ndp)$.}\label{fig:simu_ad}
	\end{figure}
	
	\section{Applications}\label{sec:app}
	In this section, we will demonstrate the proposed Wasserstein gradients for time-dependent household income and human mortality data. As mentioned before, the underlying densities are practically never known and need to be estimated from data that they generate. In the household income and mortality examples, the data are reported in the form of histograms, respectively life tables. 
	Our methods can be applied in a straightforward way to histogram data; specifically we estimate the densities by applying a smoothing step, e.g., using local linear regression.  
	For local Fr\'echet regression, we use the Epanechnikov kernel function $\ker(\tm) = 0.75(1-\tm^2)\mbf{1}_{[-1,1]}(\tm)$ and choose smoothing bandwidths $\bw$ by cross validation.
	
	\subsection{Household Income Data}\label{sec:app_inc}
	Many studies have been conducted on income distribution and inequality \citep{jone:97,heat:10}, since this is a major measure of economic equality/inequality. The evolution of income distributions over time is of particular interest as it provides quantification of the directions in which income inequality is evolving. The US Census Bureau provides histogram data of US household income over calendar years from 1994 to 2016, available at \url{https://census.gov}. 
	To make incomes of different years comparable, adjustments for inflation have been made, using the year 2000 as baseline for constant dollars.
	
	\begin{figure}[hbt!]
		\centering
		\includegraphics[width=.85\textwidth]{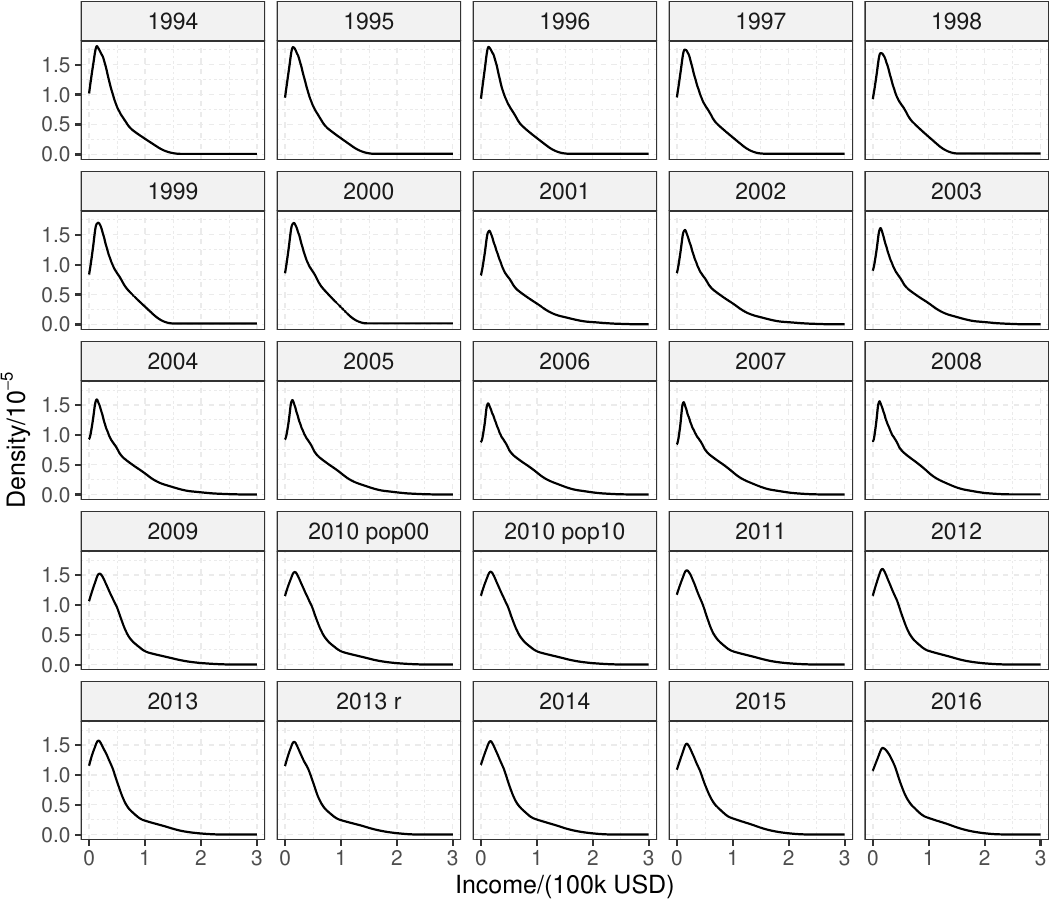}
		\caption{Densities of US household incomes for 1994--2016, where ``2010 pop00'' and  ``2010 pop10'' represent the distribution of 2010 based on the population census of 2000 and 2010, respectively, and  ``2013'' and ``2013 r'' represent the distributions on previous and redesigned questionnaires, respectively.}\label{fig:inc_data}
	\end{figure}
	
	We focus on incomes less than \$$300,000$. The data require some preprocessing, as the width of the histogram bins changed between 2000 and 2001; for 2010, due to census changes, two datasets are available based on both census 2010 and 2000 populations; for 2013, two sets of data are also available and one of them is based on a redesigned questionnaire which has been used since then. To mitigate against these changes, which potentially introduce artificial variation, we divided the whole period into four parts: 1994--2000, 2001--2010, 2010--2013 and 2013--2016. Although another change of bin width occurred between 2008 and 2009, we keep the entire period 2001--2010 in order to cover the financial crisis of 2008 well within the time interval. The densities constructed by smoothing the histogram data are shown in Figure~\ref{fig:inc_data}, where ``2010 pop00'', ``2010 pop10'', ``2013'' and ``2013 r'' represents the income distribution of 2010 based on the population census of 2000 and 2010, and of 2013 based the previous and redesigned questionnaires, respectively. 
	
	Figure~\ref{fig:inc_data} reveals not much variation in the income distributions over time except around 2008. The estimated Wasserstein temporal gradients as per \eqref{eq:wtgEst} (with bandwidths \update{$\bw=1.99$, 1.55, 1.50 and 1.50} years for the four periods, respectively, chosen by cross validation, see Section~\ref{sec:simu} and time increment $\tmincEst=\bw$) demonstrate how the income of poor, middle-class and rich households evolved  for the four periods in Figure~\ref{fig:inc_wtg}. \update{The Wasserstein temporal gradients for the ending years of each period cannot be well estimated due to the relative large value of $\tmincEst$, and hence the results for 2000, ``2010 pop00'', ``2013'' and 2016 are not displayed.} 
	Since the local Fr\'echet regression has increased variance near endpoints, the estimated Wasserstein temporal gradients on the two ends of each period are somewhat unreliable.
	
	\begin{figure}[hbt!]
		\centering
		\begin{subfigure}[c]{\textwidth}
			\centering
			\includegraphics[width=.53\linewidth]{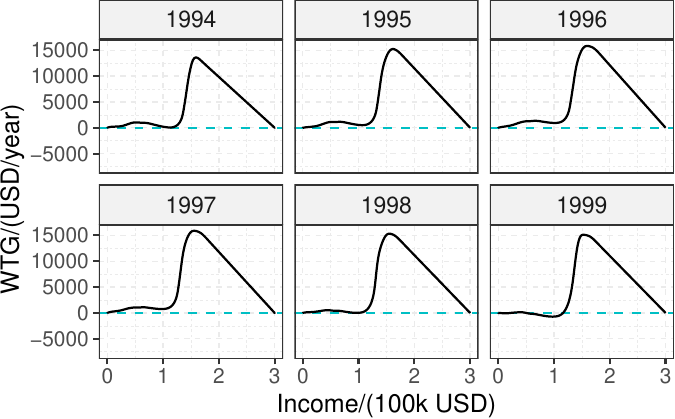}
			\caption{1994--1999.}
		\end{subfigure}\vspace{1em}
		\begin{subfigure}[c]{\textwidth}
			\centering
			\includegraphics[width=.85\linewidth]{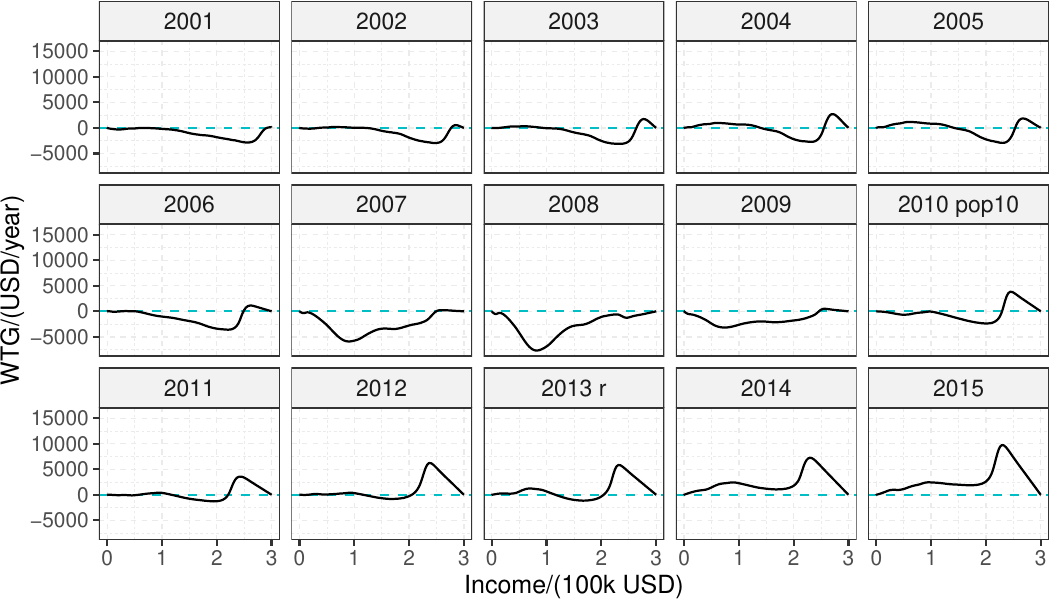}
			\caption{2001--2015.}
		\end{subfigure}
		\caption{Estimates of the Wasserstein temporal gradients (solid curves) as per \eqref{eq:wtgEst} for US household income distribution flows for 1994--1999 and 2001--2015.  Positive values indicate an increasing trend; negative values indicate a decreasing trend. 
		}\label{fig:inc_wtg}
	\end{figure}
	
	It can be seen in Figure~\ref{fig:inc_wtg} that for the period 1994--1999,  incomes of households at the same percentile levels  increased almost throughout, except for relatively poor households whose incomes tended to decrease in 1999. 
	Incomes of households earning more than \$150,000 per year increased much faster than the other incomes. 
	For the second period 2001--2010, the economic status of the lower and middle earners was stable in the first three years, rose in 2004--2006, and then declined starting in 2007. 
	Higher incomes declined  until 2002, and beginning in 2003, a divide manifested itself in the higher income levels: 
	The lower tier of higher incomes was associated with declining income, whereas the higher tier was associated with increasing income, except for \update{2007 and 2008. Note that in 2007 and 2008}, all household incomes tended to decrease, coinciding with the financial crisis. 
	For the last two periods, it can be seen that household incomes gradually recovered from the crisis. While top incomes above 240,000 US dollars always gained, households with relatively low incomes did not recover until \update{around} 2014.
	
	\subsection{Human Mortality Data}
	The analysis of mortality data across countries and species has found interest in demography and statistics \citep{care:92,chio:09,ouel:11,hynd:13,shan:17}. Of particular interest is how the distribution of age-of-death evolves over time. 
	The Human Mortality Database (\url{http://www.mortality.org}) provides data of yearly life tables for 37 countries, from which the distributions of ages-at-death in terms of histograms can be extracted. 
	
	We focus on ages-at-death in the age interval $[0,100]$ (in years) and take Russia, Sweden and the United States as three examples. The densities obtained by smoothing the histogram data for females and males separately are shown in Figures~\ref{fig:mort_russia}, \ref{fig:mort_sweden} and \ref{fig:mort_us}, respectively. 
	It can been seen that densities of mortality and their changes vary across these three countries, which is partly due to the different domains in terms of calendar years during which country-specific mortality has been recorded, which goes much further into the past for Sweden than for the other countries. 
	Estimates of the Wasserstein temporal gradients have been obtained with bandwidths $\bw$ chosen by cross validation as discussed in Section~\ref{sec:simu} per gender and country (see Table~\ref{tab:mort_bw} for details) and \update{$\tmincEst = \bw$}.
	
	\begin{table}[hbt!]
		\centering
		\caption{Bandwidths used in the local Fr\'echet regression for the age-at-death distributions.}\label{tab:mort_bw}
		\begin{tabular}{lccc}
			\toprule
			& Russia & Sweden & USA \\ 
			\midrule
			Females & 1.87 & 2.09 & 2.40 \\ 
			Males & 1.66 & 1.78 & 2.52 \\ 
			\bottomrule
		\end{tabular}
	\end{table}
	
	\begin{figure}[hbt!]
		\centering
		\includegraphics[width=.8\textwidth]{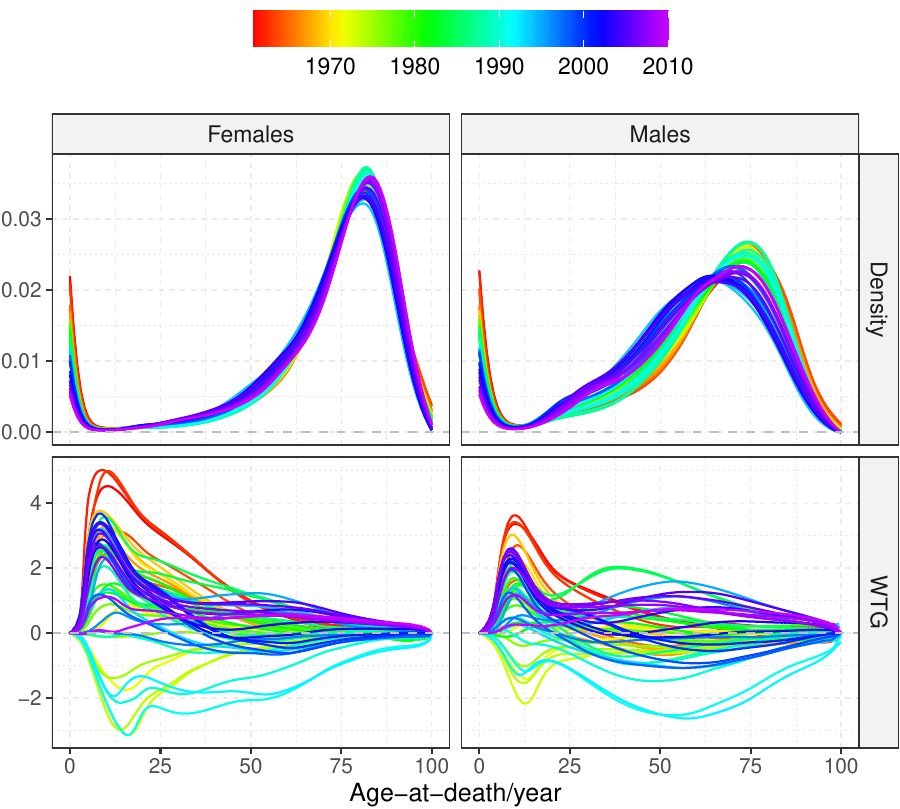}
		\caption{Top: Time-varying densities of age-at-death (in years) with females in the left column and males in the right column for Russia from \update{1961 to 2010}. Bottom: Estimates of the (unit-free) Wasserstein temporal gradients of the age-at-death (in years) distributions from \update{1961 to 2010}, where positive values indicate increasing trend and negative values indicate decreasing trend.}\label{fig:mort_russia}
	\end{figure}
	
	For Russia, the densities of ages-at-death from \update{1961 to 2010} are shown in the top two panels in Figure~\ref{fig:mort_russia}. The age-at-death distributions are quite different between females and males; female adults tend  to live longer than males. 
	The estimates of the Wasserstein temporal gradients for \update{1961--2010} for Russia are shown in the bottom two panels in Figure~\ref{fig:mort_russia}, and were obtained based on data from 1959 to 2014; the estimated gradients for the first and last two years were excluded due to boundary effects. 
	Between 1970 and 2000, the movement of mortality to higher ages and thus longer life was interrupted, with a lot of variation during this period, and resumed only in the 2000s, where substantial improvement occurs in children's mortality. 
	In the 1990s, there was a remarkable reversal in the trend of longevity increase, as the estimates of the Wasserstein temporal gradients were  negative for those years, especially for young females and mid-age males. 
	
	\begin{figure}[hbt!]
		\centering
		\includegraphics[width=.8\textwidth]{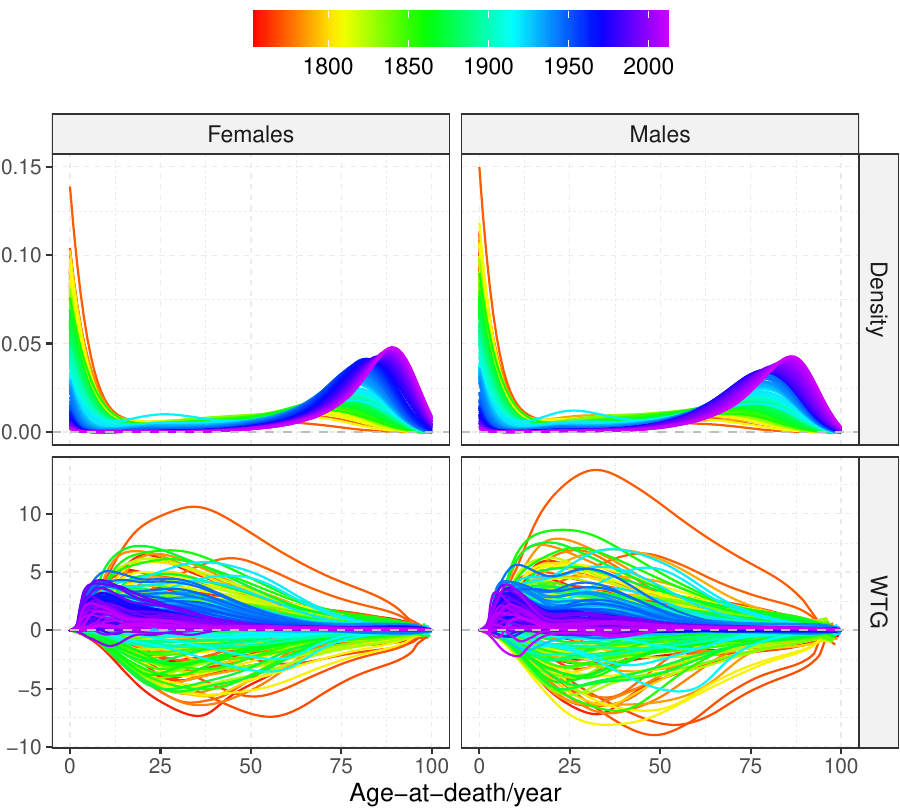}
		\caption{Top: Time-varying densities of age-at-death (in years) with females in the left column and males in the right column for Sweden from \update{1754 to 2012}. Bottom: Estimates of the (unit-free) Wasserstein temporal gradients for the distributions of age-at-death (in years) distributions from \update{1754 to 2012}, where positive values indicate increasing trend; negative values indicate decreasing trend.}\label{fig:mort_sweden}
	\end{figure}
	
	For Sweden, as shown in Figure~\ref{fig:mort_sweden}, the densities of ages-at-death of females and males are quite similar, indicating a general increase in longevity over the years. 
	The estimated Wasserstein temporal gradients for Sweden in Figure~\ref{fig:mort_sweden} from \update{1754 to 2012, which are obtained based on data from 1751 to 2016,} show some volatility in the age-at-death distributions for both females and males, especially before 1950. 
	Compared to Russia, the evolution of the age-at-death distributions in Sweden is more balanced---years where the distribution moves to the left (right) are followed by years with a rightward (leftward) movement in the distribution. 
	The Wasserstein temporal gradients for Sweden vary in a much larger range than Russia, which is partly due to the inclusion of early calendar years, where the variation of mortality from year to year was much larger, compared to more recent calendar years. 
	For example, the top orange curve for \update{1773} demonstrates a massive increasing trend in life span for both females and males while the bottom orange curves for \update{1770--1771} demonstrate a strongly decreasing trend. 
	
	For the US, the age-at-death distributions are somewhat similar across genders. The estimates of the Wasserstein temporal gradients from \update{1936 to 2010 obtained based on data from 1933 to 2015} for the US in Figure~\ref{fig:mort_us_wtg} indicate that age-at-death distributions tend to move to the right in almost all years, suggesting increasing longevity. However, for several of the years since the 1980s, reversals can be found for both females and males. A major reversal can be found for the males from young adults to middle age during \update{1983--1987}. This puzzling reversal has been attributed to drug use \citep[e.g.,][]{case:15}.
	
	\begin{figure}[hbt!]
		\centering
		\includegraphics[width=.8\textwidth]{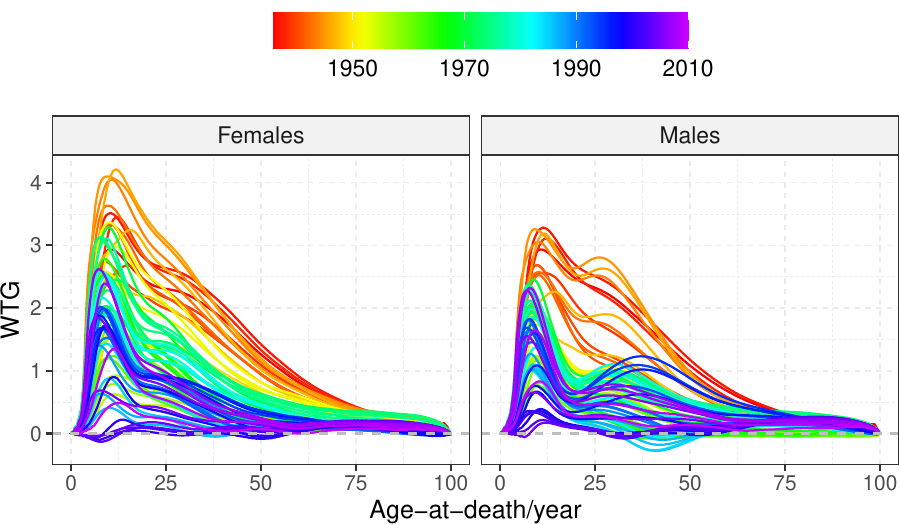}
		\caption{Estimates of the (unit-free) Wasserstein temporal gradients of the age-at-death (in years) distributions for the US from \update{1936 to 2010}, with females on the left and males on the right. Positive values indicate increasing trend; negative values indicate decreasing trend.}\label{fig:mort_us_wtg}
	\end{figure}
	
	\Appendix
	\setcounter{equation}{0}
	\renewcommand{\theequation}{A.\arabic{equation}}
	\section[Derivation of Theorem~1]{\update{Derivation of Theorem~\ref{thm:wtgEst}}}
	For $\tm\in\tdom$, we define 
	\bal\nn
	\wqTrue_{\tm}(\cdot) &= \expect\left[\wLocDf{\rtm}\qtdtn(\cdot)\right],\\
	\wqEmp_{\tm}(\cdot) &= \nobj\inv\sum_{\objidx=1}^{\nobj} \wLocDf{\tmi}\qtdtni(\cdot),\\
	\wqOrc_{\tm}(\cdot) &= \nobj\inv\sum_{\objidx=1}^{\nobj} \wLocEstDf{\tmi}\qtdtni(\cdot),\\ 
	\wqEst_{\tm}(\cdot) &= \nobj\inv\sum_{\objidx=1}^{\nobj} \wLocEstDf{\tmi}\qtdtniEst(\cdot), \eal
	where $\qtdtn$, $\qtdtni$ and $\qtdtniEst$ are the quantile functions of $\dtn$, $\dtni$ and $\dtniEst$, respectively. 
	Considering any fixed $\tm\in\tdom\interior$, we will show that \note{if $(\partial^3/\partial\tm^3)\qtCm(\parg)$ is continuous function of $\tm$ such that $\sup_{\parg\in(0,1)}|(\partial^3/\partial\tm^3)\qtCm(\parg)|<\infty$, and $\denRtm(\tm)>0$ and $|\denRtm'(\tm)|<\infty$,}
	\bal\label{eq:rate_dq_wqtrue}
	\left\ltwoNorm{\frac{\wqTrue_{\tm+\bw}-\wqTrue_{\tm}}{\bw} - \frac{\partial\qtCm}{\partial\tm}\right} = \O(\bw).\eal
	For $\wqTrue_{\tm}$, we will show that
	with sufficiently small $\bw$, \bal\label{eq:wqTrueInc}
	\partial\wqTrue_{\tm}(\parg)/\partial\parg>0,\quad \text{for all }\parg\in(0,1),\eal
	\note{if $\sup_{\tm\in\tdom}|\partial^3\qtCm(\parg)/(\partial\tm^2\partial\parg)| < \infty$ and $\sup_{\rarg\in\dom}\denCm(\rarg)<\infty$,} i.e., $\wqTrue_{\tm}$ is a quantile function, whence we will show that 
	\bal\label{eq:wqcontrol}
	\left\ltwoNorm{\qtLmEst - \wqTrue_{\tm}\right} \le \left\ltwoNorm{\wqEst_{\tm} - \wqTrue_{\tm}\right}.\eal
	Hence, 
	\bal\nn
	&\left\ltwoNorm{\frac{\qtLmEst[\tm+\bw]-\qtLmEst}{\bw} - \frac{\wqTrue_{\tm+\bw}-\wqTrue_{\tm}}{\bw}\right} \\
	&\quad\le \bw\inv \left(\left\ltwoNorm{\qtLmEst[\tm+\bw] - \wqTrue_{\tm+\bw}\right} +  \left\ltwoNorm{\qtLmEst - \wqTrue_{\tm}\right} \right)\\
	&\quad\le \bw\inv \left(\left\ltwoNorm{\wqEst_{\tm+\bw} - \wqTrue_{\tm+\bw}\right} +  \left\ltwoNorm{\wqEst_{\tm} - \wqTrue_{\tm}\right} \right). \eal
	Furthermore, we will show that 
	\bal\label{eq:rate_wqdiff}
	\left\ltwoNorm{\wqEst_{\tm+\bw} - \wqOrc_{\tm+\bw}\right} &= \Op\left((\drate\bw)\half\right),\\
	\left\ltwoNorm{\wqOrc_{\tm+\bw} - \wqEmp_{\tm+\bw}\right} &= \Op\left((\nobj\bw)\mhf\right),\\
	\left\ltwoNorm{\wqEmp_{\tm+\bw} - \wqTrue_{\tm+\bw}\right} &= \Op\left((\nobj\bw)\mhf\right). \eal
	Similar results hold when replacing $\tm+\bw$ with $\tm-\bw$. We note that for all $\tm\in\tdom$, $\ltwoNorm{\qtLmVEst-\qtLmEst}\le \bwAtom$, a.s. In conjunction with the atomlessness of $\dtnCm$ and $\dtnLmVEst$, taking $\tmincEst=\bw$ yields
	\bal\nn
	&\left\ltwoNorm{\ptr{\dtnLmVEst}{\dtnCm}\wtgVEst - \wtg\right}_{\dtnCm}\\
	&\quad=\left\ltwoNorm{\frac{\qtLmVEst[\tm+\bw]-\qtLmVEst}{\bw} - \frac{\partial\qtCm}{\partial\tm}\right}\\
	&\quad= \O(\revtwo{\bw}) + \Op\left((\nobj\bw^3)\mhf\right) + \O(\bwAtom\bw\inv) + \Op\left((\drate\bw\inv)\half\right),\eal
	whence \eqref{eq:wtg_rate} follows with \revtwo{$\bw\sim\nobj^{-1/5}$, $\bwAtom=\O(\nobj^{-2/5})$ and $\drate=\O(\nobj^{-3/5})$}. 
	Next, we will prove \eqref{eq:rate_dq_wqtrue}--\eqref{eq:rate_wqdiff}, respectively. 
	
	For any given $\tmcenter\in\tdom\interior$, there exists $\radUnif>0$ such that $\titv\subset\tdom$. 
	We note that under \ref{ass:ker} and \ref{ass:jointdtn}, as $\bw\ra 0$, it holds for $\momidx\in\{0,1\}$ that \note{(requiring kernels with compact support)}
	\bal\label{eq:kmomRate}
	&\expect\left[\kerh(\rtm - \tm)(\rtm - \tm)^{2\momidx}\right]
	= \bw^{2\momidx} \left[\denRtm(\tm) \kermom{1}{2\momidx} + \O(\bw^2)\right],\\
	&\expect\left[\kerh(\rtm - \tm)(\rtm - \tm)^{2\momidx+1}\right]
	= \bw^{2\momidx+2} \left[\denRtm'(\tm) \kermom{1}{2\momidx+2} + \o(1)\right],\\
	&\expect\left[\kerh(\rtm - \tm)|\rtm-\tm|^{2\momidx+1}\right] = \O(\bw^{2\momidx+1}), \\
	&\expect\left[\kerh(\rtm - \tm) (\rtm - \tm)^{\momidx}\right]^2 
	= \O(\bw^{2\momidx-1}), \eal 
	where $\kermom{k}{l} = \int_{-1}^1 \ker(\vartm)^k \vartm^l \diffop\vartm$, for $k,l\in\ntnum$ and the $\O$ and $\o$ terms are uniform in $\tm\in\titv$. For the following proofs, we define  $\kmom_{\momidx}^+(\tm) = \expect[\kerh(\tmi-\tm)|\tmi-\tm|^{\momidx}]$ and $\kmomEst_{\momidx}^+(\tm) = \nobj\inv\sumn [\kerh(\tmi-\tm)|\tmi-\tm|^{\momidx}]$, for $\momidx=0,1,2,3$. 
	
	\bpf[Proof of \eqref{eq:rate_dq_wqtrue}]
	\revtwo{Applying a Taylor expansion yields
		\bal\nn
		&\left\ltwoNorm{\frac{\wqTrue_{\tm+\bw}-\wqTrue_{\tm}}{\bw} - \frac{\partial\qtCm}{\partial\tm}\right}\\
		&\quad\le \frac{1}{2\bw} \left[\expect\left|\wLoc{\rtm}{\tm+\bw}{\bw}(\rtm-\tm)^2\right| + \expect\left|\wLoc{\rtm}{\tm}{\bw}(\rtm-\tm)^2\right|\right]\\
		&\qquad \times \left[\int_0^1\sup_{\tm'\in\tdom}\left|\frac{\partial^2\qtCm[\tm'](\parg)}{\partial{\tm'}^2}\right|^2\diffop\parg\right]\half. 
		\eal
	}Under \ref{ass:ker} and \ref{ass:jointdtn}, for $\momidx=0,1,2,3$, it follows from similar arguments to \eqref{eq:kmomRate} that as $\bw\ra 0$, 
	\bal\nn
	&\expect\left|\wLoc{\rtm}{\tm+\bw}{\bw}(\rtm-(\tm+\bw))^{\momidx}\right|\\
	&\quad\le \frac{\kmom_2(\tm+\bw)\kmom_{\momidx}^+(\tm+\bw) + |\kmom_1(\tm+\bw)|\kmom_{\momidx+1}^+(\tm+\bw)}{\kvar(\tm+\bw)} = \O(\bw^{\momidx}). \eal
	Hence, 
	\bal\nn
	&\expect\left|\wLoc{\rtm}{\tm+\bw}{\bw}(\rtm-\tm)^2\right|\\
	&\quad=\expect\left|\wLoc{\rtm}{\tm+\bw}{\bw}(\rtm-(\tm+\bw))^2\right| +2\bw\expect\left|\wLoc{\rtm}{\tm+\bw}{\bw}(\rtm-(\tm+\bw))\right|\\ &\qquad +\bw^2\expect\left|\wLoc{\rtm}{\tm+\bw}{\bw}\right| \\
	&\quad=\O(\bw^2). \eal
	Similarly, $\expect|\wLoc{\rtm}{\tm}{\bw}(\rtm-\tm)^2|  = \O(\bw^2)$. 
	In conjunction with \ref{ass:dtnCm}, \eqref{eq:rate_dq_wqtrue} follows. 
	\epf
	
	\bpf[Proof of \eqref{eq:wqTrueInc}]
	We note that by \eqref{eq:kmomRate} and \ref{ass:dtnCm}, 
	\bal\nn
	&\sup_{\parg\in(0,1)}\left|\frac{\partial\wqTrue_{\tm}(\parg)}{\partial\parg} - \frac{\partial\qtCm(\parg)}{\partial\parg}\right| \\
	&\quad\le \expect\left|\wLocDf{\rtm}(\rtm-\tm)\right| \sup_{\tm'\in\tdom,\,\parg\in(0,1)} \left|\frac{\partial^2\qtCm[\tm'](\parg)}{\partial{\tm'}\partial\parg}\right|\\
	&\quad= \O(\bw),\eal
	and that 
	\bal\nn
	\inf_{\parg\in(0,1)}\frac{\partial\qtCm(\parg)}{\partial\parg} 
	=\left(\sup_{\rarg\in\dom}\denCm(\rarg)\right)\inv 
	\ge C\inv >0. \eal
	Thus, with sufficiently small $\bw$, $\partial\wqTrue_{\tm}(\parg)/\partial\parg > 0$, for all $\parg\in(0,1)$. 
	\epf
	
	\bpf[Proof of \eqref{eq:wqcontrol}]
	\bal\nn
	\left\ltwoNorm{\wqEst_{\tm} - \wqTrue_{\tm}\right}^2 - \left\ltwoNorm{\qtLmEst - \wqTrue_{\tm}\right}^2 
	= \left\ltwoNorm{\wqEst_{\tm}-\qtLmEst\right}^2 + 2\left\innerprod{\wqEst_{\tm}-\qtLmEst}{\qtLmEst-\wqTrue_{\tm}\right}. \eal
	If $\innerprod{\wqEst_{\tm}-\qtLmEst}{\qtLmEst-\wqTrue_{\tm}}<0$, then there exists $\parg\in(0,1)$ such that 
	\bal\nn
	&\left\ltwoNorm{\parg\wqTrue_{\tm} + (1-\parg)\qtLmEst - \wqEst_{\tm}\right}^2 - \left\ltwoNorm{\qtLmEst - \wqEst_{\tm}\right}^2\\
	&\quad= \parg^2\left\ltwoNorm{\qtLmEst-\wqTrue_{\tm}\right}^2 + 2\parg \left\innerprod{\wqEst_{\tm}-\qtLmEst}{\qtLmEst-\wqTrue_{\tm}\right} < 0,\eal
	which contradicts the fact that $\qtLmEst = \argmin_{\qtAnyDtn} \nobj\inv\sum_{\objidx=1}^{\nobj} \wLocEstDf{\tmi} \ltwoNorm{\qtdtniEst-\qtAnyDtn}^2 = \argmin_{\qtAnyDtn}\ltwoNorm{\wqEst_{\tm}-\qtAnyDtn}^2$, where the minimization is over all the quantile functions of distributions in $\wsp$. 
	Therefore, $\innerprod{\wqEst_{\tm}-\qtLmEst}{\qtLmEst-\wqTrue_{\tm}}\ge 0$, whence \eqref{eq:wqcontrol} follows. 
	\epf
	
	\bpf[Proof of \eqref{eq:rate_wqdiff}]
	Under \ref{ass:ker}--\ref{ass:jointdtn} and \ref{ass:distn_est_rate}--\ref{ass:nObsPerDtn}, \note{(kernel is bounded)}
	\bal\nn
	&\expect\left(\left\ltwoNorm{\wqEst_{\tm+\bw} - \wqOrc_{\tm+\bw}\right}^2\right)\\
	&\le \expect\left(\frac{1}{\nobj}\sum_{\objidx=1}^{\nobj} \wLocEst{\tmi}{\tm+\bw}{\bw}^2 \left\ltwoNorm{\qtdtniEst-\qtdtni\right}^2\right)\\
	&= \expect\left[\frac{1}{\nobj}\sum_{\objidx=1}^{\nobj} \wLocEst{\tmi}{\tm+\bw}{\bw}^2 \expect\left(\left\ltwoNorm{\qtdtniEst-\qtdtni\right}^2 \cdn \tmi,\dtni\right)\right]\\
	&\le \const\drate \expect\left[\frac{1}{\nobj}\sum_{\objidx=1}^{\nobj} \wLocEst{\tmi}{\tm+\bw}{\bw}^2\right]\\
	&\le \const\frac{\drate}{\bw} \expect\left[\frac{1}{\nobj}\sum_{\objidx=1}^{\nobj} \left(\frac{\kmomEst_2(\tm+\bw)-\kmomEst_1(\tm+\bw)(\tmi-(\tm+\bw))}{\kvarEst(\tm+\bw)}\right)^2 \kerh(\tmi-(\tm+\bw))\right]\\
	&= \const\drate\bw\inv \expect\left[\frac{\kmomEst_2(\tm+\bw)^2\kmomEst_0(\tm+\bw) - \kmomEst_2(\tm+\bw)\kmomEst_1(\tm+\bw)^2}{\kmomEst_2(\tm+\bw)\kmomEst_0(\tm+\bw) - \kmomEst_1(\tm+\bw)^2}\right]\\
	&= \const\drate\bw\inv \expect\left[\kmomEst_2(\tm+\bw)\right]\\
	&= \const\frac{\drate}{\bw}\kmom_2(\tm+\bw)\\
	&\le \const\drate\bw, \eal
	whence $\ltwoNorm{\wqEst_{\tm+\bw} - \wqOrc_{\tm+\bw}} = \Op((\drate\bw)\half)$. 
	
	Next, we observe that
	\bal\nn
	\left\ltwoNorm{\wqOrc_{\tm+\bw} - \wqEmp_{\tm+\bw}\right} 
	&\le \sup_{u\in(0,1)}\frac{1}{\nobj}\sum_{\objidx=1}^{\nobj} \left|\left(\wLocEst{\tmi}{\tm+\bw}{\bw} - \wLoc{\tmi}{\tm+\bw}{\bw}\right) \qtdtni(\parg)\right|\\
	&\le \left(\max_{\rarg\in\dom}|\rarg|\right) \frac{1}{\nobj}\sum_{\objidx=1}^{\nobj} \left|\wLocEst{\tmi}{\tm+\bw}{\bw} - \wLoc{\tmi}{\tm+\bw}{\bw}\right| \\
	&\le  \left(\max_{\rarg\in\dom}|\rarg|\right) \left(|\kmomEst_0(\tm+\bw)| \left|\frac{\kmomEst_2(\tm+\bw)}{\kvarEst(\tm+\bw)} - \frac{\kmom_2(\tm+\bw)}{\kvar(\tm+\bw)} \right|\right.\\
	&\quad  + \left. |\kmomEst_1^+(\tm+\bw)| \left|\frac{\kmomEst_1(\tm+\bw)}{\kvarEst(\tm+\bw)} - \frac{\kmom_1(\tm+\bw)}{\kvar(\tm+\bw)} \right|\right). \eal
	We note that 
	\bal\label{eq:kmomBwRate} 
	\expect\left[\kerh(\rtm - (\tm+\bw)) (\rtm - (\tm+\bw))^{\momidx}\right]^2 
	= \O(\bw^{2\momidx-1}). \eal
	In conjunction with \eqref{eq:kmomRate} and Taylor expansion, 
	\bal\nn
	&\kmomEst_0(\tm+\bw) = \kmom_0(\tm+\bw) + \Op\left((\nbw)\mhf\right),\\
	&\kmomEst_1^+(\tm+\bw) = \kmom_1^+(\tm+\bw) +\Op\left(\bw(\nbw)\mhf\right),\\
	&\left|\frac{\kmomEst_2(\tm+\bw)}{\kvarEst(\tm+\bw)} - \frac{\kmom_2(\tm+\bw)}{\kvar(\tm+\bw)}\right|  =\Op\left((\nobj\bw)\mhf\right),\\ &\left|\frac{\kmomEst_1^+(\tm+\bw)}{\kvarEst(\tm+\bw)} - \frac{\kmom_1^+(\tm+\bw)}{\kvar(\tm+\bw)}\right| =\Op\left(\bw\inv(\nbw)\mhf\right), \eal
	whence we have $\ltwoNorm{\wqOrc_{\tm+\bw} - \wqEmp_{\tm+\bw}} = \Op((\nobj\bw)\mhf)$. 
	
	Lastly, we observe that
	\bal\nn
	&\left\ltwoNorm{\wqEmp_{\tm+\bw} - \wqTrue_{\tm+\bw}\right} \\
	&\quad\le \frac{\kmom_2(\tm+\bw)}{\kvar(\tm+\bw)} \frac{1}{\nobj}\sum_{\objidx=1}^{\nobj} \left\ltwoNorm{\kerh(\tmi-(\tm+\bw))\qtdtni - \expect\left[\kerh(\rtm-(\tm+\bw))\qtdtn\right]\right} \\
	&\qquad  + \frac{\bw|\kmom_1(\tm+\bw)|}{\kvar(\tm+\bw)} \frac{1}{\nobj}\sum_{\objidx=1}^{\nobj} \left\ltwoNorm{\kerh(\tmi-(\tm+\bw))\frac{\tmi-(\tm+\bw)}{\bw}\qtdtni\right. \\
		&\qquad -\left. \expect\left[\kerh(\rtm-(\tm+\bw))\frac{\rtm-(\tm+\bw)}{\bw}\qtdtn\right]\right} \\
	&\quad= \Op\left((\nobj\bw)\mhf\right), \eal
	which follows from \eqref{eq:kmomBwRate} and the boundedness of $\dom$. 
	\epf
	
	\references
\end{document}